\documentclass[pra,aps,twocolumn,showpacs,superscriptaddress]{revtex4}
\usepackage{epsfig}
\usepackage{mathrsfs}

\begin{document}

\title{Noise and deviation effects in a bichromatic Raman white light cavity}

\author{Qingqing Sun}
\affiliation{
	Department of Physics and Institute of Quantum Studies,
	Texas A\&M University,
	College Station, Texas 77843-4242
}
\author{M.\ Selim Shahriar}
\affiliation{
	Departments of EECS and Physics and Astronomy,
	Northwestern University,
	Evanston, Illinois 60208-3118
}
\author{M.\ Suhail Zubairy}
\affiliation{
	Department of Physics and Institute of Quantum Studies,
	Texas A\&M University,
	College Station, Texas 77843-4242
}

\date{\today}

\begin{abstract}
We analyze the effects of noise and parameter deviations in a bichromatic Raman type white light cavity, with potential applications in precision measurements such as gravitational wave detection. The results show that the dispersion variation induced by parameter deviation can be controlled within $10^{-4}$. The laser phase noise decreases the dispersion magnitude while the amplitude noise increases it. Although we can always adjust the parameters to satisfy the white light condition, both noises make the cavity transmission curve uneven.
\end{abstract}


\maketitle

\newcommand{\ds}{\displaystyle}
\newcommand{\dd}{\partial}
\newcommand{\be}{\begin{equation}}
\newcommand{\ee}{\end{equation}}
\newcommand{\beq}{\begin{eqnarray}}
\newcommand{\eeq}{\end{eqnarray}}
\newcommand{\dt}{\ds\frac{\dd}{\dd t}}
\newcommand{\dz}{\ds\frac{\dd}{\dd z}}
\newcommand{\D}{\ds\left(\frac{\dd}{\dd t} + c \frac{\dd}{\dd z}\right)}

\newcommand{\w}{\omega}
\newcommand{\W}{\Omega}
\newcommand{\g}{\gamma}
\newcommand{\G}{\Gamma}
\newcommand{\E}{\hat E}
\newcommand{\s}{\sigma}
%\newcommand{\bra}{\langle}
%\newcommand{\ket}{\rangle}

%%%%%%%%%%%%%%%%%%%%%%%%%%%%%%%%%%%%%%%%%%%%%%%%%%%%%%%%%%%%%
\section{Introduction}
\label{sect:intro}  % \label{} allows reference to this section

In an Fabry-Perot cavity the round-trip phase delay is proportional to the frequency. Thus only certain discrete frequencies can be exactly resonant. If the cavity is filled with a medium which provides a negative dispersion and cancels the frequency dependence of the phase delay, a continuous range of spectrum can be resonant at the same time. Such a cavity is named as white light cavity (WLC) \cite{WLC}. For precision measurements such as gravitational wave detection \cite{Wicht,Karapetyan,Wise} and ring laser gyroscopes \cite{Shahriar}, the high sensitivity requires a high finesse Fabry-Perot cavity, at the price of a reduced bandwidth. WLC provides an effective way to increase the bandwidth and solves this dilemma.

%High finesse Fabry-Perot cavities are often used in precision measurement to improve the sensitivity, at the price of a reduced bandwidth. WLC provides an effective way to increase that bandwidth, which could be very useful in gravitational wave detection \cite{Wicht,Karapetyan,Wise} and ring laser gyroscopes \cite{Shahriar}.

The dispersion requirement for the medium is $\partial_{\nu} n = -1/ \nu$, where the refractive index $n$ is a function of the frequency $\nu$. This is the so called $\lambda$-compensation or white light condition. A lot of systems are able to provide negative dispersion with small absorption or even gain \cite{Rocco,Wicht2}. For example, for two-level atoms driven by a strong resonant field \cite{Mollow,Szymanowski}, the probe dispersion is negative around the resonance. A variation of this scheme is the degenerate two-level system \cite{Friedmann}, in which there are two degenerate ground levels. Both the resonant drive field and the probe field interact with the two transitions simultaneously. The advantage of this scheme is that it does not require a very strong drive field. For a $\Lambda$ system with a bichromatic drive field far from resonance \cite{Wang,Dogariu}, the probe field experiences a gain-doublet and the dispersion is negative at the center. Another system is double-$\Lambda$ system in which the driving field interacts with the transitions from both ground levels to one of the excite level, and the probe field interacts with the transitions from both ground levels to the other excite level \cite{Wicht}. Wicht et al. analyzed and compared these systems in detail \cite{Wicht2}. Recently Savchenkov and co-workers demonstrated white light whispering gallery mode resonators \cite{Savchenkov}. For a resonator thick enough the modal spectrum becomes essentially continuous and the high quality factor is frequency-independent.

The idea of the gain-doublet scheme is proposed by Steinberg and Chiao during their persuit of superluminal pheonomena \cite{Steinberg}. Wang et al. first realized it experimentaly in a $\Lambda$ system \cite{Wang,Dogariu}. Due to the negative dispersion, the group velocity can be superluminal or even negative. The ideal case of infinite group velocity is equivalent to the white light condition. The ability of this system to achieve the white light condition has been investigated by measuring the dispersion using a heterodyne technique \cite{Pati1}, and by measuring the transmission spectrum \cite{Pati2}. A broadband cavity response has been observed.

In order to satisfy the white light condition we need to choose the parameters carefully. However, there are always deviations from the ideal values \cite{Wicht} and statistical noise. In this paper we discuss the effects of parameter deviations and laser phase and amplitude noises in the bichromatic Raman system.

%%%%%%%%%%%%%%%%%%%%%%%%%%%%%%%%%%%%%%%%%%%%%%%%%%%%%%%%%%%%%
\section{Parameter dependence of the susceptibility}

\begin{figure}
   \begin{center}
   \begin{tabular}{c}
   \includegraphics[height=5cm,width=0.9\columnwidth]{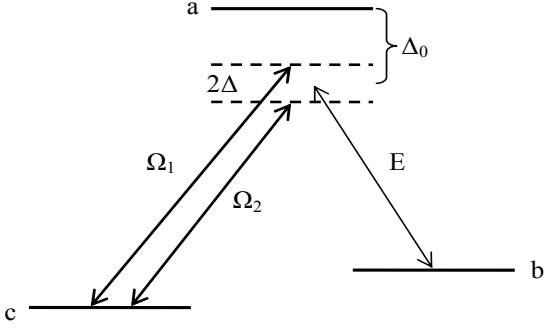}
   \end{tabular}
   \end{center}
   \caption[double Raman] 
   { \label{fig:double Raman} 
The scheme of the bichromatic Raman system. The pump fields are far detuned from the single photon transition $\left|a\right\rangle \leftrightarrow \left|c\right\rangle$ and provides a gain-doublet for the probe field. }
   \end{figure}

The level structure of the bichromatic Raman system is shown in Fig.~\ref{fig:double Raman}. There are two drive fields with frequencies $\nu_1$ and $\nu_2$ and Rabi frequencies $\Omega_1$ and $\Omega_2$, respectively. They are far detuned from the transition $\left|a\right\rangle \leftrightarrow \left|c\right\rangle$ with the detunings $\Delta_{0}+\Delta$ and $\Delta_{0}-\Delta$, where $\Delta = (\nu_1 -\nu_2)/2$ and $\Delta_0 = \omega_{ac} - (\nu_1 + \nu_2)/2$. The probe field frequency $\nu$ scans across the two Raman transitions. Such a gain doublet provides the negative dispersion at the center.

%-------------
%-------------

\begin{figure}
   \begin{center}
   \begin{tabular}{c}
   \includegraphics[height=5cm,width=0.9\columnwidth]{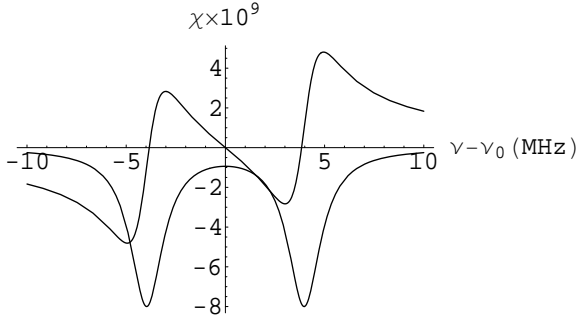}
   \end{tabular}
   \end{center}
   \caption[susceptibility] 
   { \label{fig:susceptibility} 
The probe susceptibility of a typical bichromatic Raman system. The symmetric curve is the imaginary part and the anti-symmetric curve is the real part. The dispersion inside the gain-doublet is negative.}
   \end{figure}

The susceptibility of the probe field can be written as \cite{Dogariu}
\be
\chi(\nu) = \frac{M_{1}}{(\nu-\nu_{0}-\Delta)+i\Gamma} + \frac{M_{2}}{(\nu-\nu_{0}+\Delta)+i\Gamma},
\label{Eq:OriginalSus}
\ee
where $\nu_{0} = \frac{1}{2} (\nu_{1} + \nu_{2}) - \omega_{bc}$ is the probe central frequency, $M_{j} = N (|\mu_{ab} \cdot \hat{e}|^{2}/2 \hbar \epsilon_{0}) (|\Omega_{j}|^{2}/\Delta^{2}_{0}), (j = 1, 2)$, and $\Gamma$ is the Raman transition line broadening. Usually we have $M_{1} \cong M_{2} = M$ to get the symmetrical gain peaks. Typical susceptibility curves are shown in Fig.\ref{fig:susceptibility}.

%-------------
%-------------

From the susceptibility, we can determine the refractive index $n$ and the absorption coefficient $\alpha$. The resulting expressions at the central frequency are
\be
n \cong 1 + \frac{1}{2} \chi' = 1 + \frac{1}{2} \frac{(-M_{1}+M_{2}) \Delta}{\Delta^{2}+\Gamma^{2}},
\ee

\be
\alpha \cong \frac{\nu_{0}}{2c} \chi'' = - \frac{\nu_{0}}{2c} \frac{(M_{1}+M_{2}) \Gamma}{\Delta^{2}+\Gamma^{2}},
\label{Eq:absorption}
\ee
where $\chi'$ and  $\chi''$ are the real and imaginary parts of the susceptibility $\chi$. The dispersion at $\nu_{0}$ is given by
\be
\partial_{\nu} n = - \frac{M_{1}+M_{2}}{2} \frac{(\Delta^{2}-\Gamma^{2})}{(\Delta^{2}+\Gamma^{2})^{2}}.
\label{Eq:dispersion}
\ee
By choosing the parameters carefully we can have a dispersion equal to $- 1/\nu_{0}$. The white light condition is then satisfied.

% of the pump field intensity $I_j$ (proportional to $|\Omega_j|^2$) and the number density $N$ on the white light character

In order to analyze the effect of the parameter deviations, we note that $M_{j}$ is proportional to both the pump field intensity $I_j$ and the number density $N$. Therefore the deviations of $I_j$ or $N$ lead to the variation of the absorption, dispersion and refractive index as
\be
\frac{\delta \left( \partial_{\nu} n \right)}{\partial_{\nu} n} = \frac{\delta \alpha}{\alpha} = \frac{\delta M_{1} + \delta M_{2}}{2M},
\ee

\be
\delta n = \frac{1}{2} \frac{(- \delta M_{1} + \delta M_{2}) \Delta}{\Delta^{2}+\Gamma^{2}}.
\label{Eq:Deltan}
\ee

The two intensity deviations can be independent from each other. From the proportionality between $M_{j}$ and $I_{j}$ we get
\be
\frac{\delta \left( \partial_{\nu} n \right)}{\partial_{\nu} n} = \frac{\delta \alpha}{\alpha} = \frac{1}{2} (\frac{\delta I_{1}}{I_{1}} + \frac{\delta I_{2}}{I_{2}}).
\ee
It is easier to keep the white light condition if the relative intensity deviations of the two driving fields are of opposite signs.

On the other hand, number density deviation affects $M_{1}$ and $M_{2}$ simultaneously. Therefore
\be
\frac{\delta \left( \partial_{\nu} n \right)}{\partial_{\nu} n} = \frac{\delta \alpha}{\alpha} = \frac{\delta N}{N}.
\ee
From Eq.(\ref{Eq:Deltan}), the refractive index does not change under the number density deviation.

Next we consider the effect of drive frequency deviation. If the frequency $\nu_{1}$ is changed by the amount $\delta \nu_{1}$ and $\nu_{2}$ is changed by $\delta \nu_{2}$, the susceptibility would become
\be
\chi(\nu) = \frac{M}{(\nu-\nu_{0}-\Delta - \delta \nu_{1})+i\Gamma} + \frac{M}{(\nu-\nu_{0} + \Delta - \delta \nu_{2})+i\Gamma},
\ee
From the susceptibility we can derive
\be
\delta n = \frac{M (\Delta^{2} - \Gamma^{2})}{2 (\Delta^{2} + \Gamma^{2})^{2}} \left( \delta \nu_{1} + \delta \nu_{2} \right),
\ee

\be
\delta \alpha = \frac{\nu_{0} M \Delta \Gamma}{c (\Delta^{2} + \Gamma^{2})^{2}} \left( \delta \nu_{1} - \delta \nu_{2} \right),
\ee

\be
\frac{\delta (\partial_{\nu} n)}{\partial_{\nu} n} = - \frac{\Delta (\Delta^{2} - 3 \Gamma^{2})}{\Delta^{4} - \Gamma^{4}} \left( \delta \nu_{1} - \delta \nu_{2} \right).
\ee

We consider the parameters from Ref.~\cite{Pati2}, i.e., $\Delta = 3.97MHz$, $\Gamma = 1MHz$ and $\lambda = 780nm$. On substituting these values into the above expressions we obtain $\delta n = 2.07 \times 10^{-16} (s) \left( \delta \nu_{1} + \delta \nu_{2} \right)$, $\delta \alpha = 8.97 \times 10^{-10} (s/m) \left( \delta \nu_{1} - \delta \nu_{2} \right)$, and $\delta (\partial_{\nu} n) / \partial_{\nu} n = -2.05 \times 10^{-7} (s) \left( \delta \nu_{1} - \delta \nu_{2} \right)$. Compared to the double-$\Lambda$ system \cite{Wicht}, the dependence of refractive index on the drive frequency is smaller in our system. However the dependence of the dispersion and the absorption on the drive frequency are much larger. If these two drive fields are generated from the same laser then the frequency deviations would be the same and there is no dispersion and absorption variations in this case.

Based on the same argument as in Ref.~\cite{Wicht} we conclude that the variation results in bichromatic Raman type white light cavity can be controlled within $10^{-4}$. So in theory the white light cavity linewidth could be $10^{4}$ times broader than empty cavity. But of course one has to include the other imperfect effects such as the nonlinear shape of the dispersion curve, etc.

%%%%%%%%%%%%%%%%%%%%%%%%%%%%%%%%%%%%%%%%%%%%%%%%%%%%%%%%%%%%%
\section{Effect of laser phase and amplitude noise}

In the previous section, we calculated the effect of parameter deviations, or in a more strict sense, the deviation of the expectation value. Here we consider the noise effect from the drive fields. In other words, the expectation values may have satisfied the white light condition, but the random fluctuation of the laser phase and amplitude will nevertheless modify the dispersion. The phase noises account for the finite linewidth of the drive fields and the amplitude noises are responsible for the intensity fluctuations. We calculate the effect of these noise sources independentally. For simplicity sake we assume that the separation between the two Raman peaks is much larger than the Raman linewidth and therefore we can calculate the two Raman transitions independentally.

Following the expressions in Ref.~\cite{Dogariu}, the effective Hamiltonian for the system can be written as
\beq
\hat{H} &=& \hat{H_{0}} + \hat{H_{I}} = - \hbar \omega_{ab} \left| b \right\rangle \left\langle b \right| - \hbar \omega_{ac} \left| c \right\rangle \left\langle c \right|    \nonumber \\
&-& \hbar \Omega_{p} e^{-i \nu t} \left| a \right\rangle \left\langle b \right| - \hbar \Omega_{1} e^{-i \nu_{1} t} \left| a \right\rangle \left\langle c \right| + H.c.
\eeq
A usual way to account for the effect of the laser phase noise is based on density matrix equations \cite{Wodkiewicz,Sadaf}, which is convenient if the coefficient matrices commute with each other. Here we follow a somewhat different approach as the usual methods are not easily applied. In particular, we consider the state vector instead of density matrix equations. 

The state vector of the three-level atomic system is described by 
\be
\left| \psi \right\rangle = C_{a} (t) \left| a \right\rangle + C_{b} (t) e^{i \omega_{ab} t} \left| b \right\rangle + C_{c} (t) e^{i \omega_{ac} t} \left| c \right\rangle,
\ee
where $C_{a}(t)$, $C_{b}(t)$ and $C_{c}(t)$ are the slowly varying amplitudes. The equations of motion for the amplitudes of states $\left| a \right\rangle$ and $\left| b \right\rangle$ are 
\be
\dot{C_{a}} (t) = i \Omega_{1} e^{-i \Delta_{1} t} C_{c} + i \Omega_{p} e^{-i \Delta_{p} t} C_{b},
\label{Eq:Ca}
\ee
\be
\dot{C_{b}} (t) = i \Omega^{\ast}_{p} e^{i \Delta_{p} t} C_{a} - \gamma C_{b},
\label{Eq:Cb}
\ee
where $\Delta_{1} = \nu_{1} - \omega_{ac}$ is the drive field detuning, $\Delta_{p} = \nu - \omega_{ab}$ is the probe detuning, and $\gamma$ is the decay rate from level $\left| b \right\rangle$. In order to produce gain for the probe field we set the atoms to be initially in the $\left| c \right\rangle$ state. To the lowest order of approximation we can take $C_{c} \approx 1$ and $C_{b} \approx 0$. It then follows on integrating Eq. (\ref{Eq:Ca}), that 
\be
C_{a} (t) = \int^{t}_{0} i \Omega_{1} e^{ - i\Delta_{1} t'} C_{c} (t') dt'.
\ee
From Eq. (\ref{Eq:Cb}) we obtain the formal solution
\be
C_{b} (t) = \int^{t}_{0} i \Omega^{\ast}_{p} e^{ i\Delta_{p} t'} C_{a} (t') e^{-\gamma (t-t')} dt'.
\ee
The off-diagonal density matrix elemment $\rho_{ab}$ is equal to (apart from the phase factor $exp[-i\omega_{ab}t]$)
\beq
&&\left\langle C_{a} (t) C^{\ast}_{b} (t) \right\rangle = \int^{t}_{0} -i \Omega_{p} e^{ - i\Delta_{p} t'} \left\langle C_{a} (t) C^{\ast}_{a} (t') \right\rangle e^{-\gamma (t-t')} dt'      \nonumber  \\
&&= \int^{t}_{0} -i \Omega_{p} e^{-i \Delta_{p} t'} e^{- \gamma (t-t')} dt' \int^{t}_{0} i C_{c} e^{-i \Delta_{1} t''} dt'' \times                                                          \nonumber  \\
&& ~~\times \int^{t'}_{0} -i C^{\ast}_{c} e^{i \Delta_{1} t'''} \left\langle \Omega_{1} (t'')\Omega^{\ast}_{1} (t''') \right\rangle dt'''.      
\label{Eq:correlation}
\eeq

\begin{figure}
   \begin{center}
   \begin{tabular}{c}
   \includegraphics[height=5cm,width=0.9\columnwidth]{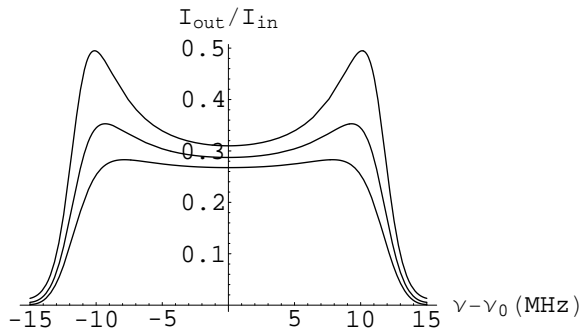}
   \end{tabular}
   \end{center}
   \caption[transmission] 
   { \label{fig:transmission} 
The transmission of the white light cavity. White light condition is satisfied in all curves. The only difference is that the decay rate increases from lower curve to higher curve.}
   \end{figure}

In order to consider the effect of phase noise, we can write the drive Rabi frequency as $\Omega_{1} (t) = \Omega_{1} e^{i \phi_{1} (t)}$. As well known the phase fluctuation of a laser is a Wiener-Levy process, i.e., the random phase with Gaussian statistics performs a Brownian motion.
\beq
\left\langle \phi_{1} (t) \right\rangle &=& 0,   \nonumber \\
\left\langle \phi_{1} (t) \phi_{1} (t') \right\rangle &=& D_{1} (t + t' - \left| t - t' \right|),
\eeq
where $D_{1}$ is the phase induced bandwidth. This gives us the correlation
\be
\left\langle \Omega_{1} (t)\Omega^{\ast}_{1} (t') \right\rangle = \left| \Omega_{1} \right|^{2} \left\langle e^{i\phi_{1}(t) - i\phi_{1}(t')} \right\rangle = \left| \Omega_{1} \right|^{2} e^{-D_{1} |t-t'|}.
\label{Eq:Wiener-Levy}
\ee
On subsituting from Eq. (\ref{Eq:Wiener-Levy}) into Eq.(\ref{Eq:correlation}) we obtain
\beq
\left\langle C_{a} (t) C^{\ast}_{b} (t) \right\rangle &\cong& \frac{\Omega_{p} |\Omega_{1}|^{2}}{\Delta^{2}_{0}} \frac{e^{-i \Delta_{p} t}}{(\Delta_{p}-\Delta_{1}) + i(\gamma + D_{1})}                                                \nonumber  \\
&& +{\rm other~frequencies}.
\eeq
In the last step we used the far detuned condition $\Delta_{0} \approx \Delta_{1} \approx \Delta_{p} >> D_{1}, \gamma$ to ignore the small terms. There are also some terms with other frequencies which do not contribute to the probe susceptibility. We recall that the polarization $P = N \mu_{ab} \rho_{ab} = \chi \epsilon_{0} E_{p}$ where the population matrix element $\rho_{ab} = \left\langle C_{a} (t) C^{\ast}_{b} (t) \right\rangle e^{-i \omega_{ab} t}$. Therefore with both Raman transitions the probe susceptibility under phase noises is
\be
\chi(\Delta_{p}) = \frac{M_{1}}{(\Delta_{p}-\Delta_{1}) + i(\gamma + D_{1})} + \frac{M_{2}}{(\Delta_{p}-\Delta_{2}) + i(\gamma + D_{2})}.
\ee
This is Eq. (\ref{Eq:OriginalSus}) if we take $\Gamma_{j} = \gamma + D_{j}, (j = 1, 2)$. The inclusion of phase noise effectively increases the width of the gain peaks. From Eqs. (\ref{Eq:dispersion}) we find that larger $\Gamma$ decreases the magnitude of the dispersion. In order to keep the white light condition we can adjust the parameters. For example, we can increase the field intensity to get larger $M_{j}$, or alternatively we can use a smaller $\Delta$. Although the dispersion condition can be restored, there is still an impact to the cavity transmission, as shown in Fig. (\ref{fig:transmission}). All the three curves are under white light condition with the same parameters except that $\Gamma$ increases from the lowest to the highest curve. We find that the transmission bandwidth is slightly increased but the curve becomes more uneven, which is not favorable.

%-------------
%-------------

Next we consider the effect of amplitude noise, $\Omega_{1} (t) = \Omega_{1} + \delta \Omega_{1} (t)$. The Gaussian type fluctuation can be described by an Ornstein-Uhlenbeck stochastic process \cite{amplitudenoise}
\beq
\left\langle \delta \Omega_{1} (t) \right\rangle &=& 0,                                                                   \nonumber \\
\left\langle \delta \Omega_{1} (t) \delta \Omega_{1} (t') \right\rangle &=& I_{\Omega1} A_{1} e^{-A_{1} |t-t'|},
\label{Eq:Ornstein-Uhlenbeck}
\eeq
where $I_{\Omega1}$ is the variance of amplitude fluctuations and $A_{1}$ is the amplitude fluctuation induced bandwidth. Again by subsituting from Eq. (\ref{Eq:Ornstein-Uhlenbeck}) into Eq.(\ref{Eq:correlation}) we obtain
\begin{widetext}
\beq
\left\langle C_{a} (t) C^{\ast}_{b} (t) \right\rangle \cong  \frac{\Omega_{p} |\Omega_{1}|^{2}}{\Delta^{2}_{0}} \frac{e^{-i \Delta_{p} t}}{(\Delta_{p}-\Delta_{1}) + i\gamma} + \frac{\Omega_{p} I_{\Omega1} A_{1}}{\Delta^{2}_{0}} \frac{e^{-i \Delta_{p} t}}{(\Delta_{p}-\Delta_{1}) + i(\gamma + A_{1})}   + {\rm other~frequencies},
\label{Eq:amplitudenoise}
\eeq
\beq
\chi(\Delta_{p}) = \frac{M_{1}}{(\Delta_{p}-\Delta_{1}) + i\gamma} + \frac{I_{\Omega1} A_{1}}{|\Omega_{1}|^{2}} \frac{M_{1}}{(\Delta_{p}-\Delta_{1}) + i(\gamma + A_{1})}
+ \frac{M_{2}}{(\Delta_{p}-\Delta_{2}) + i\gamma} + \frac{I_{\Omega2} A_{2}}{|\Omega_{2}|^{2}} \frac{M_{2}}{(\Delta_{p}-\Delta_{2}) + i(\gamma + A_{2})}.
\label{Eq:susceptibility}
\eeq
\end{widetext}
Similarly we have ignored the small terms in Eq. (\ref{Eq:amplitudenoise}). In Eq. (\ref{Eq:susceptibility}) both Raman transitions are included to find the susceptibility under amplitude noise. The net effect are the two additional terms which are similar to the original terms with only different coefficients and $\gamma$ changed to $\gamma + A_{j}$. Therefore both the dispersion and gain will increase in magnitudes. Still we can satisfy the white light condition by adjusting the parameters, for example we can decrease the drive field intensity. Similarly we will find the cavity transmission curve becomes uneven since the two additional terms have a larger linewidth $\gamma + A_{j}$.

%%%%%%%%%%%%%%%%%%%%%%%%%%%%%%%%%%%%%%%%%%%%%%%%%%%%%%%%%%%%%
\section{Conclusion}
In this paper we consider the impact of parameter deviations and laser phase and amplitude noises on a bichromatic Raman type white light cavity. We find the dispersion, which needs to satisfy the white light condition, can be controlled within $10^{-4}$ under the parameter deviations. Therefore a white light cavity could have $10^{4}$ times broader linewidth compared to an empty cavity at the same finesse.

The phase noise effectively increases the Raman linewidth by the diffusion $D$, causing a smaller dispersion. The amplitude noise introduces an additional term in the probe susceptibility and makes the dispersion larger. These opposite effects allow us to easily adjust the parameters to satisfy the white light condition. Both noises have the effect of making the transmission curve uneven for the white light cavity.

%%%%%%%%%%%%%%%%%%%%%%%%%%%%%%%%%%%%%%%%%%%%%%%%%%%%%%%%%%%%%
\begin{acknowledgments}

This work is supported by the Qatar National Research Fund (QNRF).

\end{acknowledgments}
%%%%%%%%%%%%%%%%%%%%%%%%%%%%%%%%%%%%%%%%%%%%%%%%%%%%%%%%%%%%%

%%%%% References %%%%%


\begin{thebibliography}{99}

\frenchspacing

\bibitem{WLC} R.-H. Rinkleff and A. Wicht, Phys. Scri. ({\bf T118}), 85 (2005).

\bibitem{Wicht} A. Wicht, K. Danzmann, M. Fleischhauer, M. Scully, G. ${\rm M\ddot{u}ller}$, and R.-H. Rinkleff, Opt. Commun. {\bf 134}, 431 (1997).

\bibitem{Karapetyan} G. G. Karapetyan, Opt. Commun. {\bf 238}, 35 (2004).

\bibitem{Wise} S. Wise, G. Mueller, D. Reitze, D. B. Tanner, and B. F. Whiting, Class. Quantum Grav. {\bf 21}, S1031 (2004).

\bibitem{Shahriar} M. S. Shahriar, G. S. Pati, R. Tripathi, V. Gopal, M. Messall, and K. Salit, Phys. Rev. A {\bf 75}, 053807 (2007).

%\bibitem{Fleischhauer} M. Fleischhauer, C. H. Keitel, M. O. Scully, C. Su, B. T. Ulrich, and S.-Y. Zhu, Phys. Rev. A {\bf 46}, 1468 (1992).

\bibitem{Rocco} A. Rocco, A. Wicht, R.-H. Rinkleff, and K. Danzmann, Phys. Rev. A {\bf 66}, 053804 (2002).

\bibitem{Wicht2} A. Wicht, R.-H. Rinkleff, L. S. Molella, and K. Danzmann, Phys. Rev. A {\bf 66}, 063815 (2002).

\bibitem{Mollow} B. R. Mollow, Phys. Rev. {\bf 188}, 1969 (1969).

\bibitem{Szymanowski} K. Szymanowski, A. Wicht, and K. Danzmann, J. Mod. Opt. {\bf 44}, 1373 (1997).

\bibitem{Friedmann} H. Friedmann and A. D. Wilson-Gordon, Opt. Commun. {\bf 98}, 303 (1993).

%\bibitem{Akulshin} A. M. Akulshin, S. Barreiro, and A. Lezama, Phys. Rev. Lett. {\bf 83}, 4377 (1999).

\bibitem{Wang} L. J. Wang, A. Kuzmich, and A. Dogariu, Nature {\bf 406}, 277 (2000).

\bibitem{Dogariu} A. Dogariu, A. Kuzmich, and L. J. Wang, Phys. Rev. A {\bf 63}, 053806 (2001).

\bibitem{Savchenkov} A. A. Savchenkov, A. B. Matsko, and L. Maleki, Opt. Lett. {\bf 31}, 92 (2006).

\bibitem{Steinberg} A. M. Steinberg and R. Y. Chiao, Phys. Rev. A {\bf 49}, 2071 (1994).

\bibitem{Pati1} G. S. Pati, R. Tripathi, M. Messall, K. Salit, and M. S. Shahriar, e-print arXiv: quant-ph/0512260.

\bibitem{Pati2} G. S. Pati, M. Salit, K. Salit, and M. S. Shahriar, Phys. Rev. Lett. {\bf 99}, 133601 (2007).

\bibitem{Wodkiewicz} K. Wodkiewicz and M. S. Zubairy, Phys. Rev. A {\bf 27}, 2003 (1983).

\bibitem{Sadaf} S. Sultana and M. S. Zubairy, Phys. Rev. A {\bf 49}, 438 (1994).

\bibitem{amplitudenoise} Selected Papers on Noise and Stochastic Process, edited by N. Wax (Dover, New York, 1954).

\end{thebibliography}
\end{document}